\documentclass[a4paper,11pt]{article}
\usepackage{pos}

\title{Quarkonium in non-zero isospin chemical potential environment at $T \simeq 0$}

\author*[a]{Seyong Kim}
\author[b]{Bastian B. Brandt}
\author[c]{Gergely Endr\H{o}di}

\affiliation[a]{Department of Physics, Sejong University, \\
Seoul 05006, Republic of Korea}

\affiliation[b]{Fakult\"at f\"ur Physik, Universit\"at Bielefeld, \\
D-33615 Bielefeld, Germany}

\affiliation[c]{Institute of Physics and Astronomy, E\"otv\"os Lor\'and University,\\
H-1117 Budapest, Hungary}

\emailAdd{skim@sejong.ac.kr}
\emailAdd{brandt@physik.uni-bielefeld.de}
\emailAdd{gergely.endrodi@ttk.elte.hu}

\abstract{We study how the isospin asymmetry affects quarkonium states in QCD at near zero temperature. Using lattice Non-Relativistic QCD formalism, we calculate bottom quark correlators in the gauge field ensembles generated with $N_f = 2 + 1$ flavors of dynamical staggered quarks whose dynamics include the isospin chemical potential effect and then construct $S-$ and $P-$ wave quarkonium state correlators. From these quarkonium correlators, we consider the ratios of quarkonium correlators at non-zero isospin chemical potential to that at $\mu_I a = 0.000$. Here, the gauge field ensemble with $\mu_I a = 0.000, 0.048, 0.053, 0.059, 0.066, 0.080, 0.092$ and $0.106$ on a $32^3 \times 48$ lattice with non-zero isospin current strength $\lambda a = 0.0010, 0.0018,$ and $0.0036$, where $m_\pi = 135$ MeV and $a = 0.1535$ fm from \cite{Brandt:2022hwy}, are used. Preliminary results suggest that for $\mu_I a = 0.106$, the Upsilon mass gets heavier than the Upsilon mass in the vacuum and that below $\mu_I a = 0.106$ the isospin asymmetry effect on the Upsilon mass is not monotonic.}

\FullConference{The 42nd International Symposium on Lattice Field Theory (LATTICE2025)\\
2-8 November 2025\\
Tata Institute of Fundamental Research, Mumbai, India\\}

\begin{document}
\maketitle

\section{Introduction}

Quantitative understanding on the part of the QCD phase diagram at non-zero baryon density  is important for understanding many physical phenomena involving many-body system of hadrons: upcoming experiment of Compressed Baryonic Matter (CBM) at FAIR will produce wealthy of information on the highly compressed baryonic matter and thus theoretical understanding is needed to disentangle various aspects of QCD in the experimental data. Also, since the Equation of State (EoS) of QCD matter at high baryon densities plays a significant role for the neutron star structure \cite{Tolman:1939jz,Oppenheimer:1939ne} and for neutron star mergers \cite{abbott2017gw170817}, accurate astrophysical observations in the future will require a better understanding of QCD at high baryon densities. QCD at non-zero baryon densities may leave a gravitational wave "footprint" even in the QCD epoch of early universe \cite{Franciolini:2023wjm}. 

\begin{figure}[hbt]
\centering
\includegraphics[width=0.4\textwidth]{./Figures/S-wave_j0.02-0.04m5.0.eps}
\includegraphics[width=0.43\textwidth]{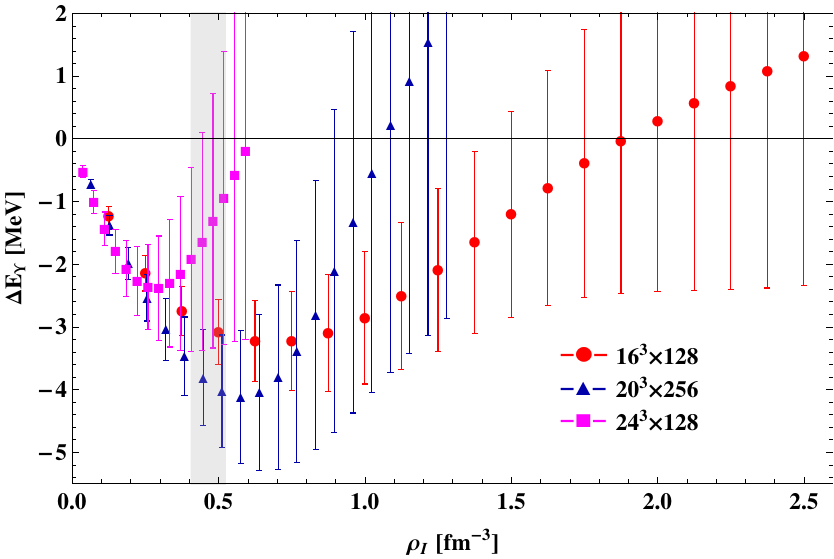}
\caption{The change in the mass of $^1S_0$ and $^3S_1$ quarkonium states in the $SU(2)$ gauge theory vs. the baryon chemical potential (the left) \cite{Hands:2012yy}. For QCD with an isospin asymmetry, the change in the mass of $\Upsilon$ state vs. the isospin charge density(the right) \cite{Detmold:2012pi}.}
\label{fig:otherworks}
\end{figure}

However, theoretical understanding of QCD at high baryon densities is hampered by the non-perturbative nature of QCD at low energy and "complex action" problem. Lattice QCD simulations which make a significant contribution to quantitative understanding of QCD matter at various temperatures have met difficulties in simulating QCD with non-zero baryon chemical potential other than the small region of QCD phase diagram near $\mu_B \sim 0$. Despite the numerous efforts, this hardship seems to be difficult to overcome \cite{deForcrand:2009zkb}. In contrast, QCD-like system such as $SU(2)$ gauge theory and $G_2$ gauge theory at high "baryon" densities, and QCD with isospin chemical potential do not have the complex-action problem and are amenable to Monte Carlo simulations. Although these systems are different from QCD at high baryon densities, lessons learned from these many-body systems may help us to understand QCD at non-zero baryon density \cite{Kogut:2000ek}. This aspect is noted early on  \cite{Son:2000by,Son:2000xc,Splittorff:2000mm,Kogut:2002zg,Kogut:2002tm,Kogut:2004zg}, and prompted many studies on QCD with isospin asymmetry \cite{Beane:2007es,Detmold:2008fn,Detmold:2012pi,Brandt:2017oyy,Brandt:2022hwy,Abbott:2023coj}. Interestingly, the EoS from the phase-quenched version of QCD can put a bound on the EoS of QCD matter at high baryon densities \cite{Cohen:2003ut,Cohen:2003kd,Moore:2023glb,Abbott:2024vhj}. For the case of zero strange quark chemical potential, the phase-quenched version of QCD with quark chemical potential exactly coincides with QCD in isospin chemical potential. On the other hand compared to the EoS of QCD, dynamics of QCD at high baryon densities is more complicated and various hadronic correlators at high isospin chemical potential may reveal such a complex nature of QCD dynamics at high baryon densities.

\begin{figure}[hbt]
\centering
\includegraphics[width=0.433\textwidth]{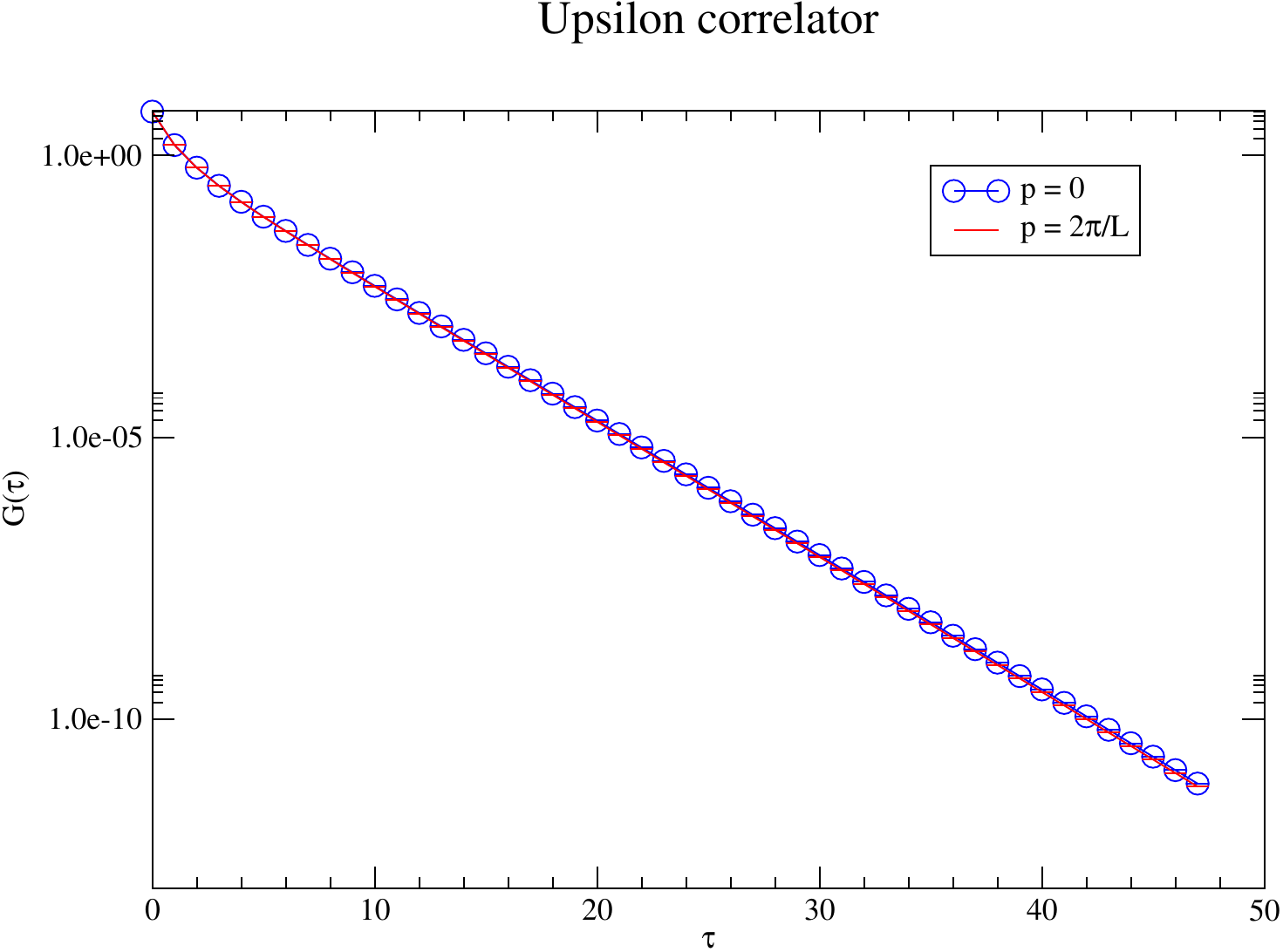} 
\includegraphics[width=0.4\textwidth]{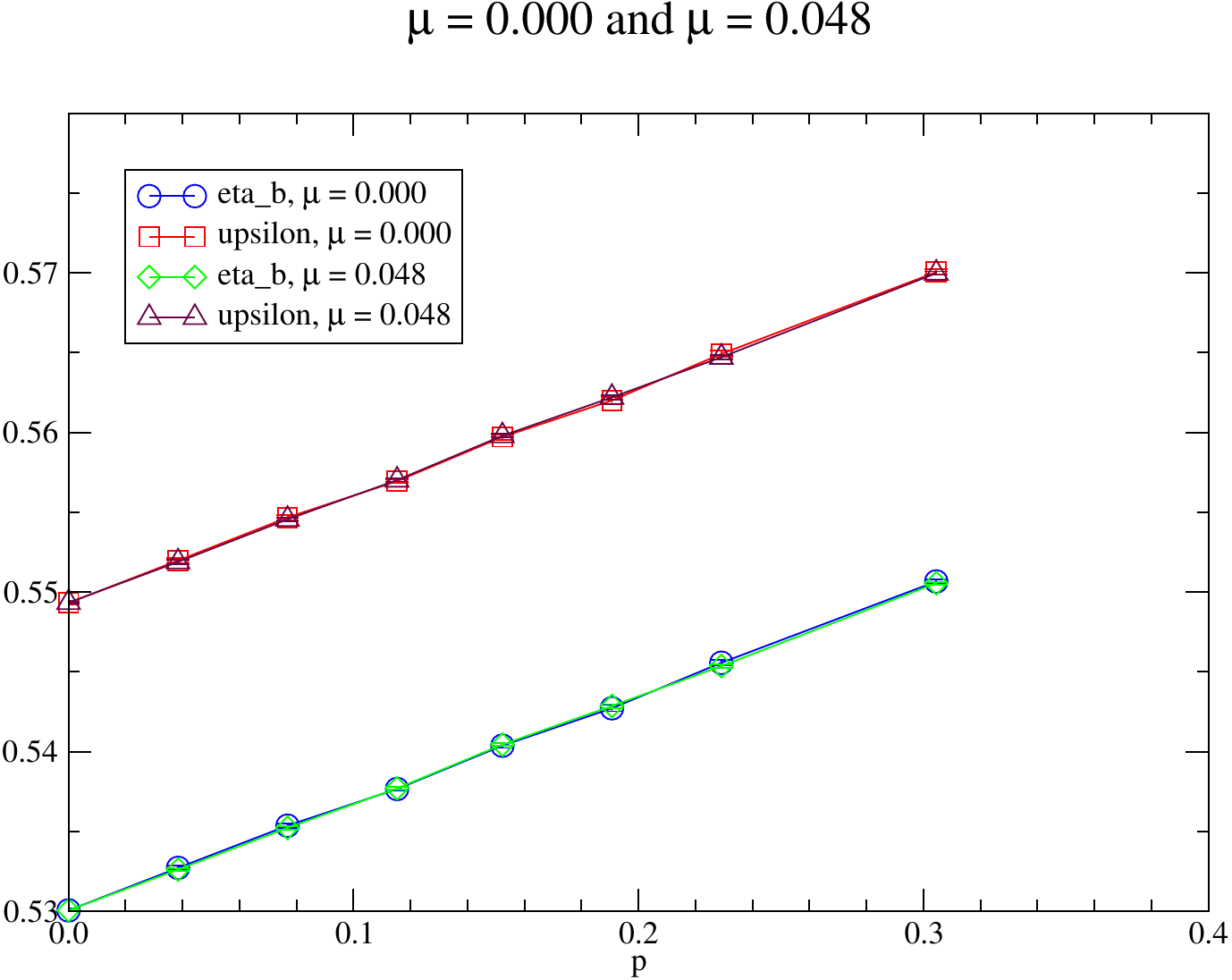}
\caption{(the left figure) Upsilon propagator with $\hat{P} = 0$ and $ \hat{P}^2 = 4 \sin^2 (\frac{2\pi}{L})$. (the right figure) $a E_{\eta_b}$ (lower line and lower data points and $a E_\Upsilon$ (upper line and upper data points) vs. lattice momentum, $\hat{P}^2$ for two different $(u,d)$ current sources, $\mu_I a = 0.0$ and $\mu_I a = 0.048$. The lines are with the fitted values from the dispersion relation.}
\label{fig:propagator}
\end{figure}

At non-zero temperature, among many hadronic states quarkonium is expected to serve as a good "thermometer" for the Quark-Gluon Plasma (QGP) \cite{Matsui:1986dk, Mocsy:2008eg} and in-medium behavior of quarkonium states at non-zero temperature are actively being pursued (see e.g., \cite{Lombardo:2025sfo,Aarts:2014cda, Kim:2018yhk, Tang:2025ypa,Ding:2025fvo}). Similarly, one hopes that the behavior of quarkonium states at high isospin densities may play the role of "densimeter" for QCD at high baryon densities. There are earlier studies on in-medium behavior of the "quarkonium states" with baryon densities for QCD-like theories: quarkonium states at non-zero baryon chemical potential in one of QCD-like theories such as $SU(2)$ gauge theory is studied in \cite{Hands:2012yy} and quarkonium in QCD with isospin charge densities is investigated in \cite{Detmold:2012pi}. There are differences in the behaviors of quarkonium states in these two theories: For $SU(2)$ gauge theory, the mass of quarkonium states at large baryon chemical potential are significantly smaller than those in the vacuum. For QCD, the mass of quarkonium states at large isospin chemical potential are larger than those in the vacuum but the change in the mass is small and is not monotonic with the isospin charge density. But whether these differences in otherwise similar theories stem from the fact that the baryon chemical potential in $SU(2)$ is different from the isospin chemical potential in $SU(3)$ QCD or from the fact that the baryon in $SU(2)$ is a di-quark state is not clear (see Fig. \ref{fig:otherworks}.). Further studies are needed.

In this proceedings, we report preliminary results from our study on in-medium behavior of the quarkonium states in isospin asymmetric QCD. The quarkonium correlators are constructed from Non-Relativistic (NR) quark correlator calculated in the background gauge fields generated with quarks whose dynamics include the isopspin chemical potential effect \cite{Brandt:2017oyy, Brandt:2022hwy}. Details of lattice setup is given in Sec. \ref{sec:setup} and preliminary results are described in Sec. \ref{sec:result}. Then, discussion on the results follows in Sec. \ref{sec:discussion}.

\section{Lattice Setup}
\label{sec:setup}
In order to investigate the properties of bottomonium in a isospin chemical potential medium, we compute the correlators of quarkonium using a discretized version of the ${\cal O} (v^4)$ NRQCD Lagrangian
\cite{Lepage:1992tx, Davies:1994mp, Gray:2005ur} for bottom quarks,

\begin{equation}
\label{LNRQCD}
{\cal L} = {\cal L}_0 + \delta {\cal L},
\end{equation}
with
\begin{equation}
\label{LNRQCD_1}
{\cal L}_0 = \psi^\dagger \left(D_\tau - \frac{{\mathbf{D}^2}}{2M_b} \right) \psi + \chi^\dagger \left(D_\tau + \frac{{\mathbf{D}^2}}{2M_b} \right) \chi,
\end{equation}
and
\begin{eqnarray}
\label{LNRQCD_2}
\delta {\cal L} = &&\hspace{-0.6cm} - \frac{c_1}{8M_b^3} \left[\psi^\dagger ({\mathbf{D}^2})^2 \psi - \chi^\dagger ({\mathbf{D}^2})^2 \chi \right] \nonumber \\
&&\hspace{-0.6cm} + c_2 \frac{ig}{8M_b^2}\left[\psi^\dagger \left({\mathbf{D}}\cdot{\mathbf{E}} - {\mathbf{E}}\cdot{\mathbf{D}}\right) \psi +
  \chi^\dagger \left({\mathbf{D}}\cdot {\mathbf{E}} - {\mathbf{E}}\cdot{\mathbf{D}} \right) \chi \right] \nonumber \\
&&\hspace{-0.6cm} - c_3 \frac{g}{8M_b^2}\left[\psi^\dagger
  \mathbf{\sigma}\cdot\left({\mathbf{D}}\times{\mathbf{E}}-{\mathbf{E}}\times{\mathbf{D}}\right)\psi + \chi^\dagger {\mathbf{\sigma}}\cdot\left({\mathbf{D}}\times{\mathbf{E}}-{\mathbf{E}}\times{\mathbf{D}}\right)\chi \right] \nonumber \\
&&\hspace{-0.6cm} - c_4 \frac{g}{2M_b} \left[\psi^\dagger {\mathbf{\sigma}}\cdot{\mathbf{B}} \psi - \chi^\dagger {\mathbf{\sigma}} \cdot {\mathbf{B}} \chi \right],
\end{eqnarray}
where $D_\tau$ and ${\mathbf{D}}$ are the gauge covariant temporal and the spatial derivatives, $\psi$ is the heavy quark and $\chi$ the heavy anti-quark. With the tadpole improvment, $c_i = 1$ are usually chosen \cite{Lepage:1992xa}.

The isospin chemical potential influences dynamics of the light $(u,d)$ quarks and the ensemble of the lattice gauge fields from \cite{Brandt:2022hwy} which include such effects is used in the computation of the lattice covariant derivatives. These ensembles are on $32^3 \times 48$ lattices with $a \simeq 0.15$fm where the pion mass is $m_\pi \simeq 135$ MeV (or $\frac{1}{2} m_\pi a = 0.053$). The $T \simeq 0$ lattice gauge field ensemble used in this preliminary work is listed in Table \ref{tab:lattice_ensemble}. Note that simulations of these isospin chemical potential ensembles are done with non-zero $(u,d)$ current sources, $\lambda$ (see \cite{Brandt:2022hwy}) to limit the small eigenvalues associated with the massless Goldstone mode of the pion condensation and $\lambda \rightarrow 0$ limit of lattice observables needs to be taken from the simulation results to remove this symmetry breaking effect.

From the discretized version of Eq. \eqref{LNRQCD}, the lattice NRQCD propagator for the bottom quark is formulated as an initial value problem
\begin{eqnarray}
G (\mathbf{x}, \tau=0) = && \hspace{-0.6cm} S(\mathbf{x}), \nonumber \\
G (\mathbf{x}, \tau=a) = && \hspace{-0.6cm} \left(1 - \frac{H_0}{2n}\right)^n
U_4^\dagger(\mathbf{x}, 0) \left(1 - \frac{H_0}{2n}\right)^n G(\mathbf{x},0), \nonumber \\
G (\mathbf{x}, \tau+a ) = && \hspace{-0.6cm} \left(1 - \frac{H_0}{2n}\right)^n
U_4^\dagger(\mathbf{x}, \tau) \left(1 - \frac{H_0}{2n}\right)^n
\left(1 -\delta H \right)  G (\mathbf{x}, \tau).\label{NRQCDEvolEq}
\end{eqnarray}
$S(\mathbf{x})$ denotes an appropriate complex valued random point source, diagonal in spin and color. This improves the signal to noise ratio by averaging. In the continuum formulation the initial condition for $G(\mathbf{x}, \tau)$ corresponds to a delta function, which we approximate on the lattice through averaging multiple correlators, started from random sources on different slices $\tau_{\rm start}$ along Euclidean time:
\begin{align}
 S^{\Upsilon, \eta_b}(\mathbf{x},\tau_{\rm start})=\eta(\mathbf{x},\tau_{\rm start}), \quad \langle \eta^\dagger(\mathbf{x}) \eta(\mathbf{x}') \rangle_{\tau_{\rm start}}=\delta_{\mathbf{x}\mathbf{x}'}
\end{align}

Here, the lowest-order Hamiltonian is 
\begin{equation}
    H_0 = - \frac{\Delta^{(2)}}{2M_b}, 
\end{equation} 
while
\begin{eqnarray}
\delta H = && \hspace{-0.6cm} - \frac{(\Delta^{(2)})^2}{8 M_b^3} + \frac{ig}{8 M_b^2}
(\mathbf{\Delta}^{\pm}\cdot \mathbf{E} - \mathbf{E}\cdot \mathbf{\Delta}^{\pm}) - \frac{g}{8 M_b^2} \mathbf{\sigma} \cdot
  (\mathbf{\Delta}^{\pm} \times \mathbf{E} - \mathbf{E}\times \mathbf{\Delta}^{\pm})   \nonumber \\
&& \hspace{-0.6cm}
- \frac{g}{2 M_b} \mathbf{\sigma}\cdot\mathbf{B}
 + \frac{a^2\Delta^{(4)}}{24 M_b} - \frac{a (\Delta^{(2)})^2}{16 n M_b^2}.
\label{eq:deltaH}
\end{eqnarray}
$n$ is the Lepage parameter, which controls effectively the temporal step size in Euclidean time and is essential to the stability of the high momentum behavior of the propagator $G$. We use $n=2$, in anticipation of the characteristic values of $M_b a$, which will arise from the tuning of $M_b a$ on the lattices used in this study, is larger than 1.

The lattice covariant derivative $\Delta$ is defined as
\begin{eqnarray}
a \Delta_i^{+} \psi (\mathbf{x}, \tau) &=& U_i (\mathbf{x}, \tau) \psi (\mathbf{x} + \hat{i} a ,
\tau) - \psi (\mathbf{x}, \tau) \nonumber \\
a \Delta_i^{-} \psi (\mathbf{x}, \tau) &=& \psi (\mathbf{x}, \tau) - U_i^\dagger (\mathbf{x}
- \hat{i} a, \tau) \psi (\mathbf{x} - \hat{i} a , \tau) \nonumber \\
\Delta^{(2)} &=& \sum_{i=1}^{3} \Delta_i^{+} \Delta_i^{-}, \;\;\;\;\;\;
\Delta^{(4)} = \sum_{i=1}^{3} (\Delta_i^{+} \Delta_i^{-})^2,
\end{eqnarray}
and the chromo-electric $({\mathbf{E}})$ and the magnetic field $({\mathbf{B}})$ are defined from clover-leaf plaquettes. The last two terms of Eq. \eqref{eq:deltaH} correct for finite lattice spacing errors. Tad pole improvement of the gauge link variable using the fourth root of a single link plaquette \cite{Lepage:1992xa} for the gauge fields in Table( \ref{tab:lattice_ensemble}) is adopted and thus $c_i$ in Eq. \eqref{LNRQCD_2} is set to the tree-level value ($c_i = 1$) after dividing the link variables by $<U_{\rm plaq}>^{1/4} \simeq 0.8510449$. 

\begin{figure}[hbt]
\centering
\includegraphics[width=0.5\textwidth]{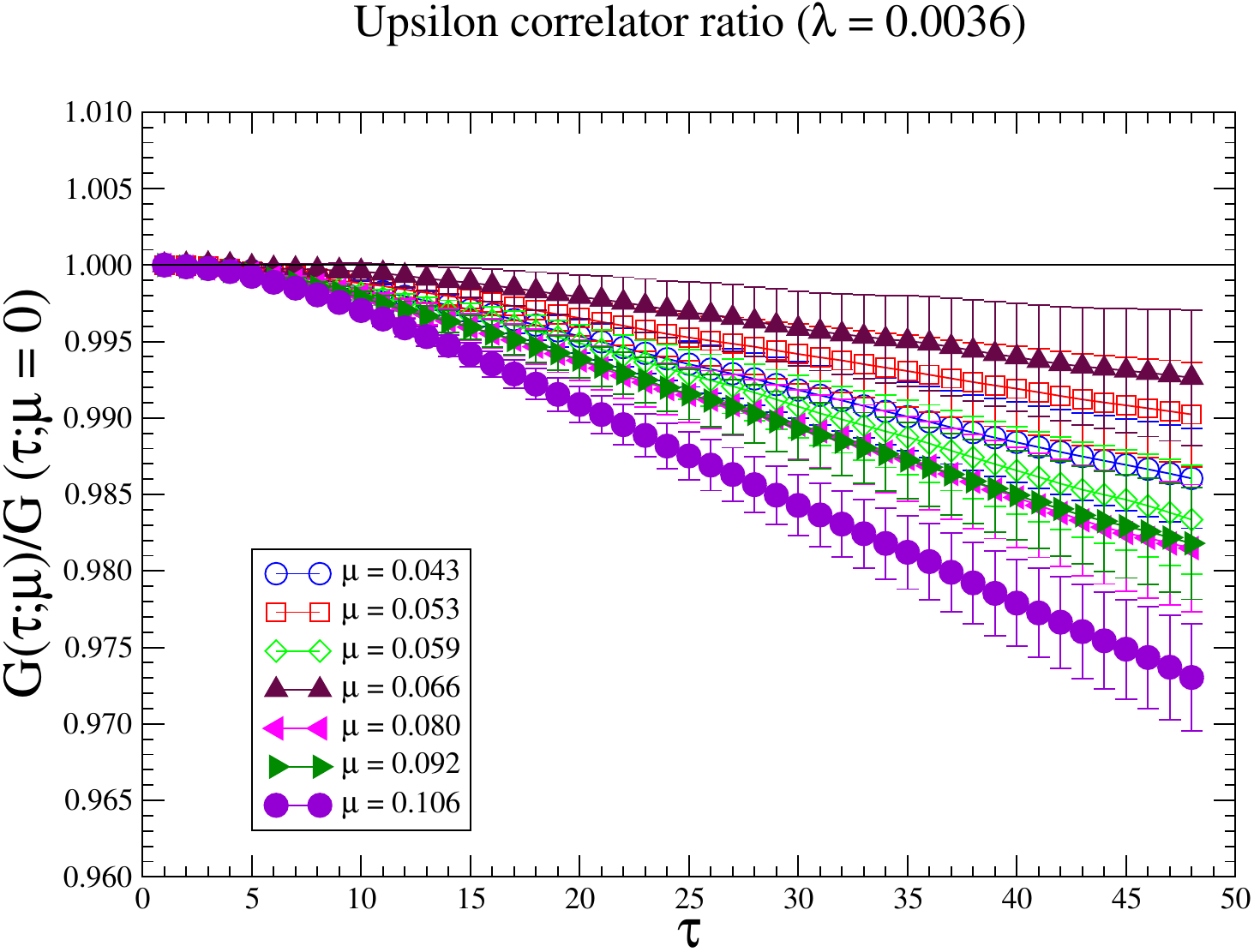} 
\caption{The ratio of the $\Upsilon$ correlators with $\lambda a = 0.0036$ for $\mu_I a = (0.000, 0.043, 0.053, 0.059, 0.066, 0.080, 0.092)$ and $0.106$ to the $\Upsilon$ correlator with $\mu_I a = 0$.}
\label{fig:fixed_j_propagator}
\end{figure}

\begin{table}[b]
\centering
\begin{tabular}{|c||c|c|c|c|c|c|c|c|}
\hline
$\mu_I a$ & 0.000 & 0.048 & 0.053 & 0.059 & 0.066 & 0.080 & 0.092 & 0.106 \\
\hline
   & 0.0010  & 0.0010 & 0.0010 & 0.0010 & 0.0010 & 0.0010 & 0.0010 & 0.0010  \\ 
$\lambda a$   & 0.0018  & 0.0018 & 0.0018 & 0.0018 & 0.0018 & 0.0018 & 0.0018 & 0.0018  \\ 
   & 0.0036  & 0.0036 & 0.0036 & 0.0036 & 0.0036 & 0.0036 & 0.0036 & 0.0036  \\ 
\hline
No. of configs. & 102 & 198 & 200 & 196 & 196 & 196 & 197 &
196 \\
\hline
\end{tabular}
\caption{The number of configurations, the magnitude of the isospin chemical potential, and the magnitude of the $(u,d)$ current sources in the ensemble used for the preliminary analysis.}
\label{tab:lattice_ensemble}
\end{table}

\section{Preliminary Result}
\label{sec:result}
To tune the bottom quark mass, we use the dispersion relation, $a E(\hat{P}^2) = a M_1 + a \frac{\hat{P}^2}{2 M_2} + \cdots $ ($\hat{P}^2 = 4 \sum_i \sin^2 (\frac{\pi n_i}{N_s})$), using the quarkonium correlators which are computed with finite lattice momenta and compare with the experimental value of the spin-averaged $1S$ upsilon mass, $\overline{M_2 (1S)} = (M_2 (\eta_b) + 3 M_2(\Upsilon))/4$ \cite{Aarts:2014cda,HPQCD:2011qwj}. With the lattice spacing of $a \simeq 0.15$ fm, $a \overline{M_2} (1S) ^{\rm exp}= 7.3480$. Fig. \ref{fig:propagator} shows an example of the Upsilon correlator (i.e., $1S$, spin triplet, $^3S_1$) behavior (the left figure) and an example of the $1S$-state energy obtained from a exponential fits to the $1S$-state (spin singlet ($\eta_b$) and spin triplet ($\Upsilon$)) correlator data for small $\hat{P}^2$'s. Here $M_b a = 3.3177$ is used. The fit to the $1S$-state energies of $\eta_b$ and $\Upsilon$ obtained from the lattice quarkonium correlators gives $a \overline{M_2}^{\rm lattice} = 7.386 (32)$, which is slightly heavier than the experimental value, $a \overline{M_2} (1S)^{\rm exp}$ within the error-bar.

To see the isospin chemical potential effect on the quarkonium correlators, the ratios of the correlators with finite $\mu_I a$ to that with $\mu_I a = 0$ are studied. Fig. \ref{fig:fixed_j_propagator} shows an example of such ratios of the upsilon correlators calculated with $\lambda a = 0.0036$ for 7 different $\mu_I a$ to the upsilon correlator calculated with $\lambda a = 0.0010$. Despite the large statistical errors, one can see the isospin chemical potential effect is small ($\ll 3 \%$ at $\tau / a = 48$). The effect can be seen as non-monotonic as a function of $\mu_I a$ from this figure. Below the pion condensation chemical potential ($\simeq \mu_I a = 0.053$), the ratios are smaller than 1 but are similar to each other. At $\mu_I a = 0.066$ (which is above the condensation chemical potential), the ratio is larger than that from $\mu_I a = 0.043$. Then the ratios from $\mu_I a = 0.080$ onward are smaller than that from $\mu_I a = 0.043$.  

\begin{figure}[hbt]
\centering
\includegraphics[width=0.3\textwidth]{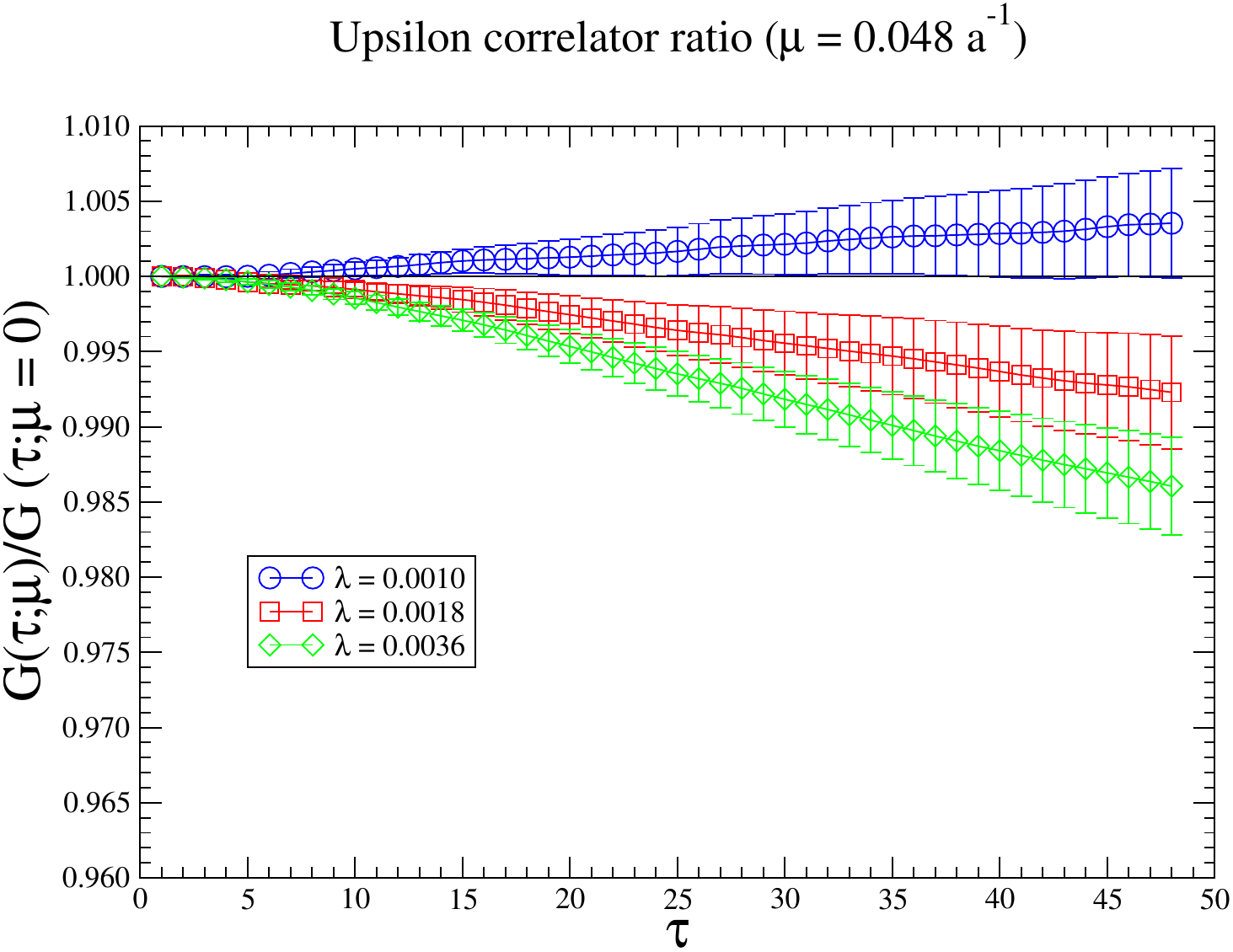}
\includegraphics[width=0.3\textwidth]{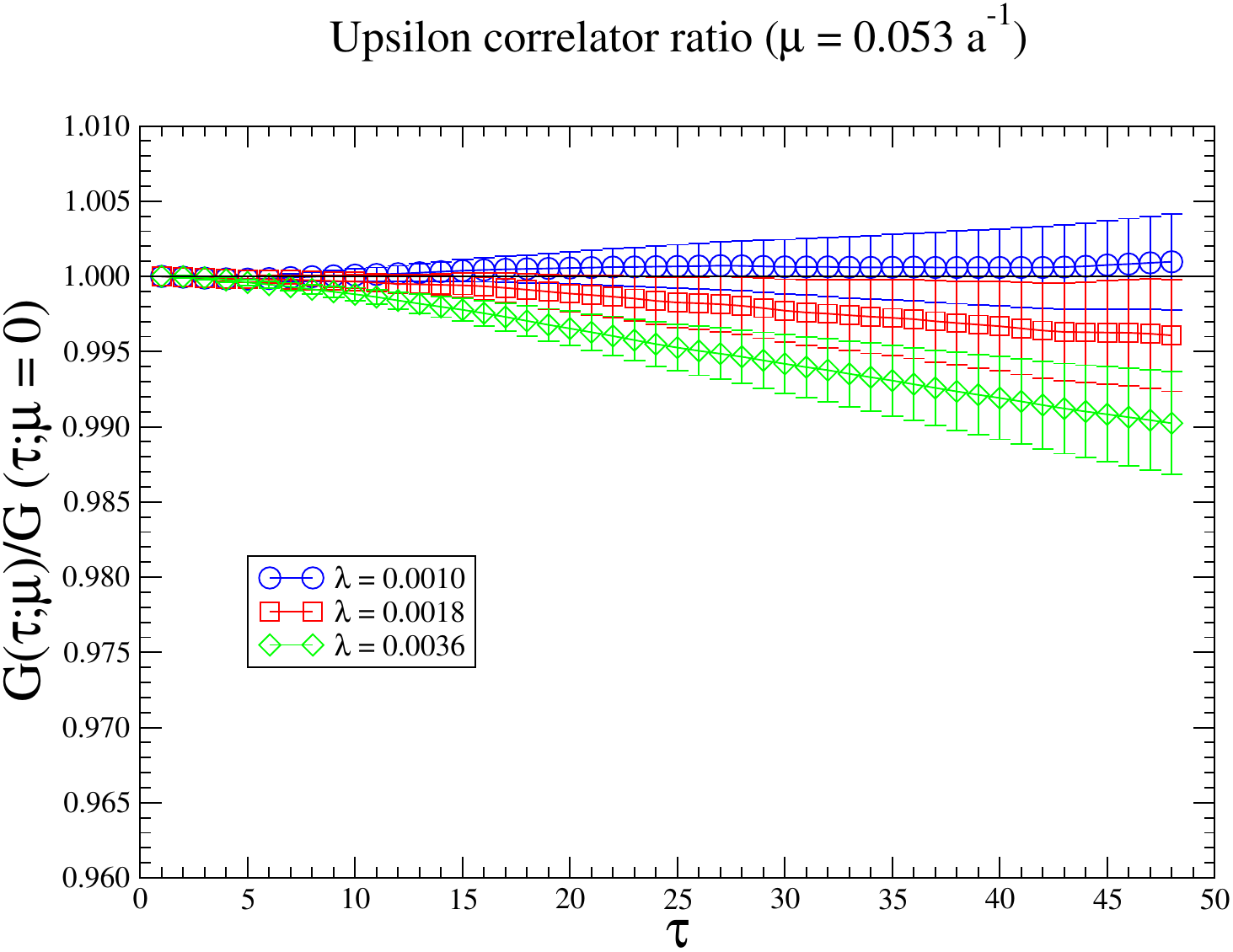}
\includegraphics[width=0.3\textwidth]{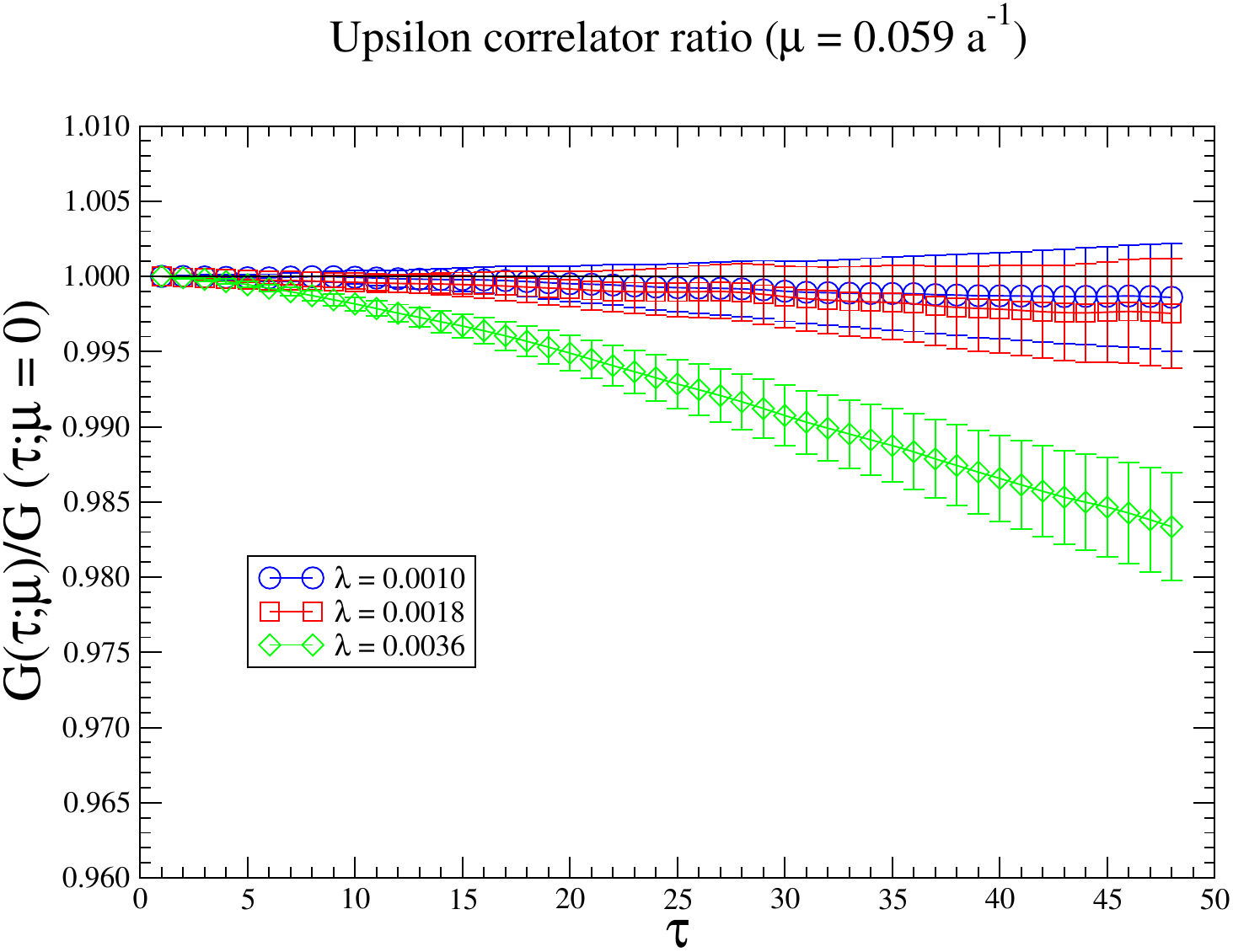}
\caption{The ratio of the $\Upsilon$ correlators with $\mu_I a = 0.048$ (the left), with $\mu_I a = 0.053$ (the center), with $\mu_I a = 0.059$ (the right) for three $\lambda a = (0.0010, 0.0018, 0.0036)$ $(u,d)$ current sources.}
\label{fig:correlator1}
\end{figure}

\begin{figure}[hbt]
\centering
\includegraphics[width=0.3\textwidth]{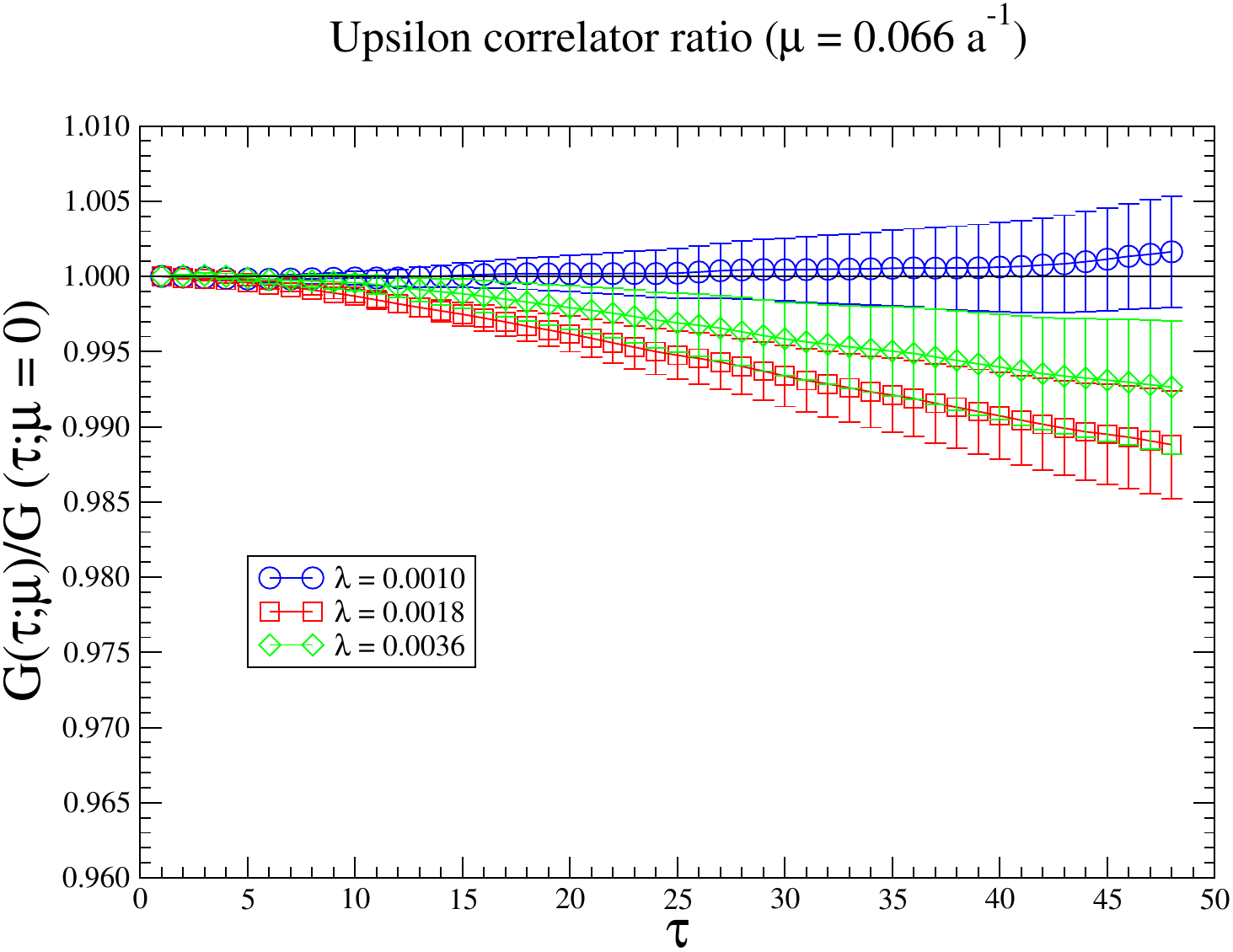}
\includegraphics[width=0.3\textwidth]{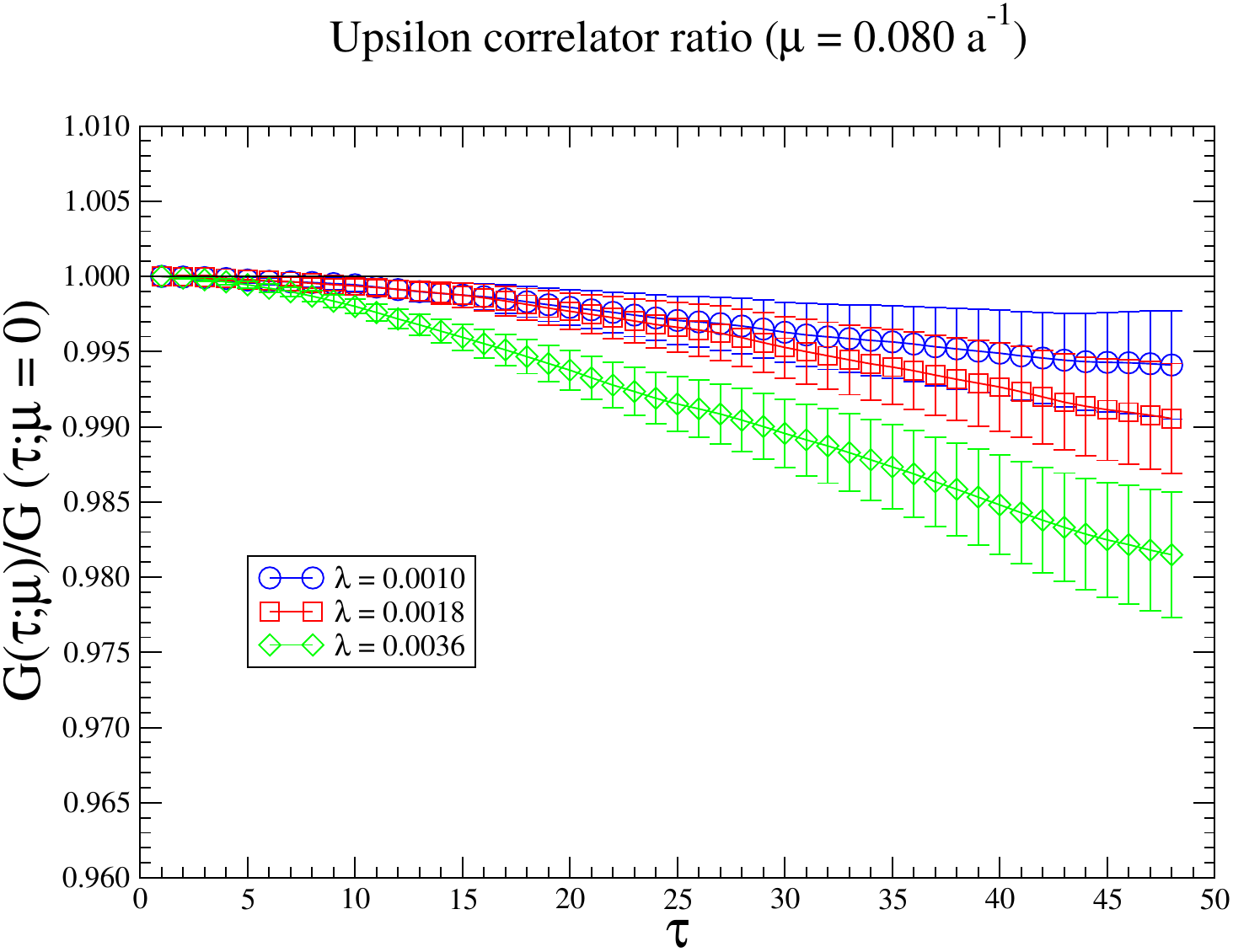}
\includegraphics[width=0.3\textwidth]{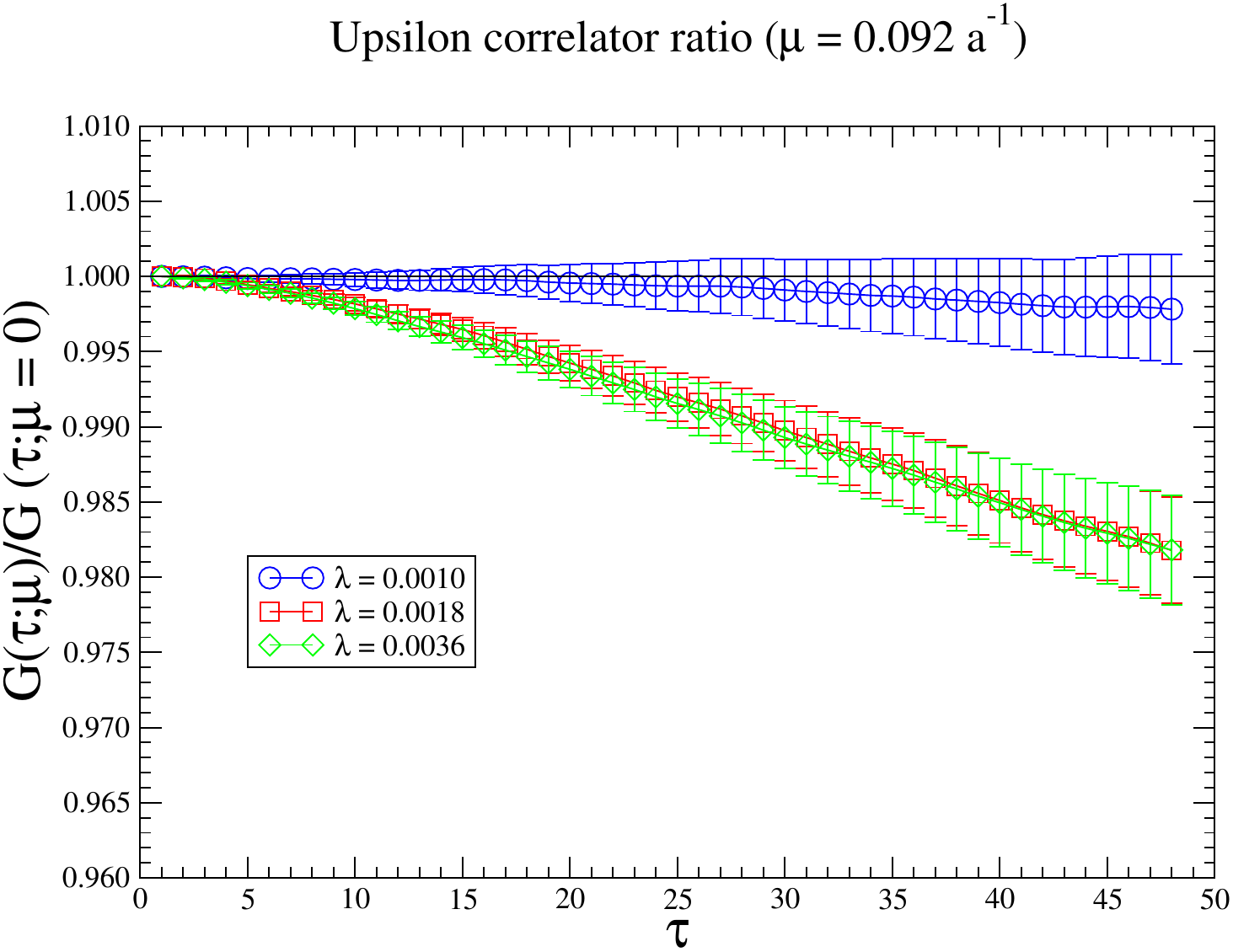}
\caption{The ratio of the $\Upsilon$ correlators with $\mu_I a = 0.066$ (the left), with $\mu_I a = 0.080$ (the center), with $\mu_I a = 0.092$ (the right) for three $\lambda a = (0.0010, 0.0018, 0.0036)$ $(u,d)$ current sources.}
\label{fig:correlator2}
\end{figure}

\begin{figure}[hbt]
\centering
\includegraphics[width=0.4\textwidth]{./Figures/upsilon_b_ratio_Mu0.048.pdf}
\includegraphics[width=0.4\textwidth]{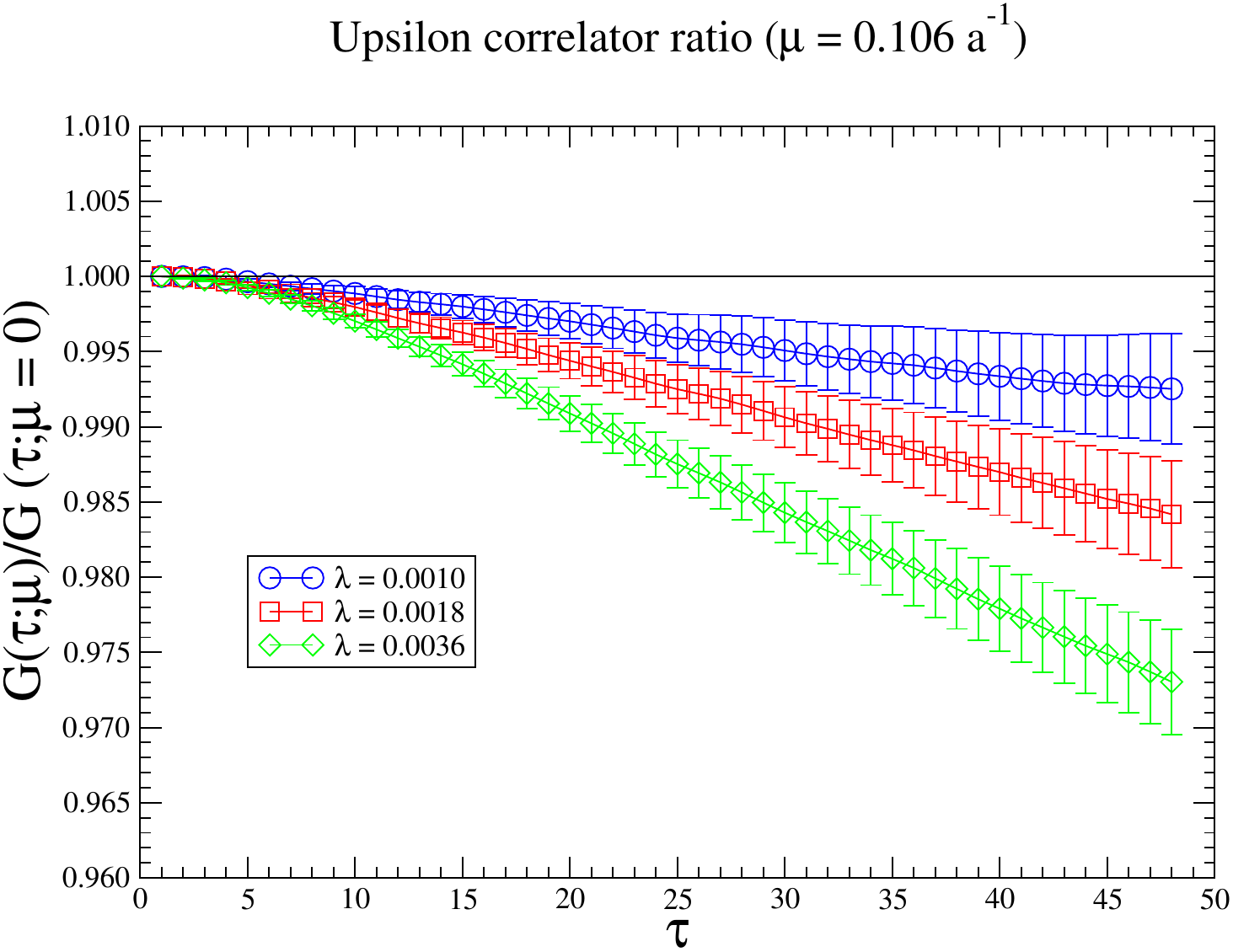}
\caption{The ratio of the $\Upsilon$ correlators with $\mu_I a = 0.048$ (the left), with $\mu_I a = 0.106$ (the right) for three $\lambda a = (0.0010, 0.0018, 0.0036)$ $(u,d)$ current sources. These two figures contrast the behavior of the correlator ratio below $\mu_I^c a = \frac{1}{2} m_\pi a \simeq 0.053$ obtained from EoS to that far above $\mu_I^c a$.}
\label{fig:correlator3}
\end{figure}

Fig. \ref{fig:correlator1}, \ref{fig:correlator2} and \ref{fig:correlator3} show the ratio of quarkonium correlators with three difference magnitudes $\lambda a = (0.0010, 0.0018, 0.0036)$ of the $(u,d)$ current source. And Fig. \ref{fig:correlator3} compare the ratios with the smallest $\mu_I a ( = 0.043)$ in the ensemble with the largest $\mu_I a ( = 0.106)$. On the other hand, larger $\lambda$ encourages the pion condensation. These comparison shows that quarkonium correlators are quite sensitive to the strength of the $(u,d)$ current source and to the isospin chemical potential. The correlator ratio of the $\mu_I a = 0.106$ case suggests a non-zero (but small) effect of the isospin chemical potential even after the $\lambda \rightarrow 0$ extrapolation. The fact that the ratio is smaller than 1 suggest that the mass of $\Upsilon$ ($^3S_1$) with the isospin chemical potential medium is heavier than that in the vaccum.

\section{Discussion}
\label{sec:discussion}
In these proceedings, we report preliminary results from our study on quarkonium at non-zero isospin chemical potential where heavy quark correlators are calculated with lattice NRQCD formulation. The $SU(3)$ gauge field ensembles \cite{Brandt:2022hwy} which include $N_f = 2+1$ light quark dynamics with an isospin asymmetry effect below and above $\mu_I \sim \frac{m_\pi}{2}$ are used.

In the isospin asymmetric medium, the mass of Upsilon state ($^3 S_1$) appears to be slightly lighter at $\mu_I a \le 0.053$ and becomes heavier for $\mu_I a \ge 0.106$ than that in the vacuum but the effects is quite small and non-monotonic with regard to increasing $\mu_I a$ since the ratio of Upsilon correlators appears to be slighter larger than 1 for $\mu_I a < 0.053$ within the statistical error and becomes smaller than 1 for $\mu_I a \ge 0.106$. The ratios depend on the strength of the $(u,d)$ current source $\lambda$ sensitively for small $\mu_I a$. For $\mu_I a = 0.048$, the ratio may become larger than 1 (the mass becomes lighter) for $\lambda a = 0.0010$ (albeit with larger error) although the ratio is smaller than 1 for $\lambda a = 0.0018$ and $\lambda a = 0.0036$. For $\mu_I a = 0.053, 0.059$ and $0.066$, the ratios for $\lambda a = 0.0010$ are consistent with 1 within the error-bar (recall that the pion condensation occurs at $\mu_I a \simeq 0.053$ \cite{Brandt:2017oyy}).

From this preliminary result, we conclude that the bottom quark mass tuning needs to be improved and that larger number of Monte Carlo samples of lattice gauge field ensembles are necessary to quantitatively study the effect of the pion condensation on the heavy quarkonium below $\mu_I a \le 0.106$. For $\mu_I a \ge 0.106$, we observe that there is a clear effect on the Upsilon state mass from the isospin chemical potential medium above the statistical uncertainty. This is in line with an earlier observation by other group which used the background of the isospin charge density to study the in-medium behavior of quarkonium \cite{Detmold:2012pi} (the right panel in Fig. \ref{fig:otherworks}). 

On the other hands the qualitative behavior of the quarkonium states in isospin chemical potential is different from the heavy quarkonium states of $SU(2)$ gauge theory at non-zero "baryon" densities where the baryonic state in $SU(2)$ gauge theory is a quark-quark bound state \cite{Hands:2012yy} (see the left panel in Fig. \ref{fig:otherworks}). The masses of $SU(2)$ theory quarkonium states at high "baryon" densities become lighter than those in the vacuum. This suggests that heavy quark bound states in QCD-like theories at high "baryon densities" may serve as a probe for the in-medium dynamics of QCD since quarkonium states discern underlying theories. We are currently increasing Monte Carlo samples of the gauge field ensembles to reduce statistical errors and are performing simulations with $\mu_I a > 0.106$ to more clearly understand isospin asymmetry effect. 

\section*{Acknowledgements}

SK is supported by the National Research Foundation of Korea under the grant NRF-2008-000458 and the grant NRF-2021-1092701, funded by the Korean government.
BB and GE acknowledge support by the Deutsche Forschungsgemeinschaft (DFG, German Research Foundation) through the CRC-TR 211 `Strong-interaction matter under extreme conditions' -- project number 315477589 -- TRR 211. GE is also supported by the Hungarian National Research, Development and Innovation Office - NKFIH (Research Grant Hungary 150241) and the European Research Council (Consolidator Grant 101125637 CoStaMM).

\bibliographystyle{JHEP}
\bibliography{Lat2025}

\end{document}